\documentclass[prl,twocolumn,floatfix,a4paper,superscriptaddress]{revtex4}
\usepackage{bm,color,graphicx,amsmath,txfonts}

\newcommand{\CC}{{\cal C}}
\newcommand{\DD}{{\cal D}}

\newcommand{\PP}{{\cal P}}

\begin{document}

\title{Magnetostrictively induced stationary entanglement between two microwave fields}

\author{Mei Yu}
\affiliation{Zhejiang Province Key Laboratory of Quantum Technology and Device, Department of Physics and State Key Laboratory of Modern Optical Instrumentation, Zhejiang University, Hangzhou, Zhejiang, China}
\author{Heng Shen}
\affiliation{Clarendon Laboratory, University of Oxford, Parks Road, Oxford, OX1 3PU, UK}
\author{Jie Li}\thanks{j.li-17@tudelft.nl}
\affiliation{Zhejiang Province Key Laboratory of Quantum Technology and Device, Department of Physics and State Key Laboratory of Modern Optical Instrumentation, Zhejiang University, Hangzhou, Zhejiang, China}
\affiliation{Kavli Institute of Nanoscience, Department of Quantum Nanoscience, Delft University of Technology, 2628CJ Delft, The Netherlands}

\begin{abstract}
We present a scheme to entangle two microwave fields by using the nonlinear magnetostrictive interaction in a ferrimagnet. The magnetostrictive interaction enables the coupling between a magnon mode (spin wave) and a mechanical mode in the ferrimagnet, and the magnon mode simultaneously couples to two microwave cavity fields via the magnetic dipole interaction. The magnon-phonon coupling is enhanced by directly driving the ferrimagnet with a strong red-detuned microwave field, and the driving photons are scattered onto two sidebands induced by the mechanical motion. We show that two cavity fields can be prepared in a stationary entangled state if they are respectively resonant with two mechanical sidebands. The present scheme illustrates a new mechanism for creating entangled states of optical fields, and enables potential applications in quantum information science and quantum tasks that require entangled microwave fields.
\end{abstract}

\date{\today}
\maketitle

Many quantum information tasks, e.g., quantum teleportation \cite{Wootters}, quantum metrology~\cite{VG}, and fundamental tests of quantum mechanics \cite{Bell}, require optical entangled states. Conventionally, they have been generated via parameteric down-conversion with nonlinear crystals~\cite{PK}. Alternative efficient approaches have been adopted by utilizing, e.g., four-wave mixing in optical fibers~\cite{fiber} and atomic vapors~\cite{atom}, quantum dots~\cite{dot}, and periodically poled lithium niobate waveguide~\cite{Gisin}, to name but a few. In the microwave (MW) domain, the entangled fields are typically produced by using the nonlinearity in Josephson parametric amplifiers~\cite{JPA}, or by injecting a squeezed vacuum through a linear MW beamsplitter~\cite{Menzel}. In the field of optomechanics, two MW fields can get entangled by coupling to a common mechanical resonator via radiation pressure~\cite{David,SB}. Despite their different forms, most of the mechanisms utilize the nonlinearity of the physical processes.

In this Letter, we present a mechanism, distinguished from all previous approaches, for creating continuous-variable (CV) entanglement of MW fields by using the nonlinear magnetostrictive interaction in a ferrimagnet. Specifically, two MW cavity fields couple to a magnon mode in a ferrimagnetic yttrium-iron-garnet (YIG) sphere~\cite{Kittel,NakaRev,Strong1,S2,S3,S4,S5,S6,S7,S8}, and simultaneously the magnon mode couples to a phonon mode embodied by the vibrations of the sphere induced by the magnetostrictive force~\cite{Tang16}. Due to the intrinsic low frequency of the phonon mode, it owns a large thermal occupation at typical cryogenic temperatures. We thus drive the magnon mode with a red-detuned MW field (with the detuning equal to the mechanical frequency), which leads to the stimulation of the anti-Stokes process, i.e., a MW photon interacts with a phonon and converts into a magnon of a higher frequency~\cite{JiePRL,JieRC}. This process corresponds to the cooling of the phonon mode, a prerequisite for observing quantum effects in the system~\cite{OMRMP}. The strong magnon drive also enhances the effective magnon-phonon coupling, and when this coupling is sufficiently strong, magnomechanical entanglement is created, similar to the mechanism of creating optomechanical entanglement with a strong red-detuned drive~\cite{DV07}. The entanglement originates from the nonlinear magnetostrictive coupling, and could be distributed to two MW fields due to the linear magnon-photon coupling. Or more intuitively, the mechanical motion scatters the MW driving photons onto two sidebands, which are entangled due to the mediation of mechanics~\cite{Genes1,Gut}. And if two MW cavities are respectively resonant with the two sidebands, the two cavity fields get entangled. Similar mechanism has been used to generate atom-light entanglement~\cite{Genes2}. We prove its validity, and two MW fields indeed get maximumly entangled when they are respectively resonant with the two mechanical sidebands (we assume the resolved sidebands~\cite{Tang16,Strong1,S2,S3,S4,S5,S6,S7,S8}). We verify the entanglement in both the quantitative and qualitative ways, i.e., by calculating the logarithmic negativity and by using the Duan criterion for CV systems. 

We first present a general model of the scheme, then solve the system dynamics by means of the standard Langevin formalism and the linearization treatment, and study the entanglement in the stationary state. Finally, we show strategies to measure/verify the optical entanglement, and provide possible configurations for experimental realizations.

\begin{figure}[t]
\includegraphics[width=\linewidth]{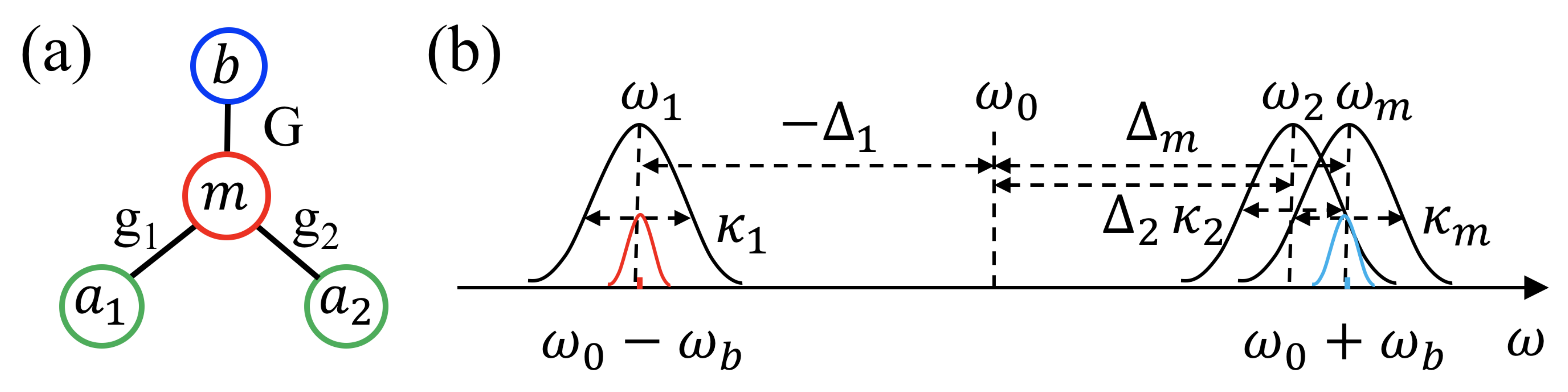}
\caption{(a) General model of the scheme. A magnon mode $m$ in a YIG sphere couples to two MW fields $a_1$ and $a_2$ via magnetic dipole interaction, and to a phonon mode $b$ via magnetostrictive interaction. (b) Mode frequencies and linewidths. The magnon mode with frequency $\omega_m$ is driven by a strong MW field at frequency $\omega_0$, and the mechanical motion of frequency $\omega_b$ scatters the driving photons onto two sidebands at $\omega_0 \pm \omega_b$. If the magnon mode is resonant with the blue (anti-Stokes) sideband, and the two cavity modes with frequencies $\omega_{1,2}$ are respectively resonant with the two sidebands, the two cavity fields get entangled. }
\label{Fig1}
\end{figure}

{\it The model.}  The system consists of two MW cavity modes, a magnon mode, and a mechanical mode, as shown in Fig.~\ref{Fig1}(a). The magnons, as quantized spin wave, are the collective excitations of a large number of spins inside a massive YIG sphere.  The magnon mode couples to two MW cavity modes via magnetic dipole interaction, and, simultaneously, to a mechanical vibrational mode via the magnetostrictive force~\cite{Tang16,JiePRL,JieRC}. The mechanical frequency we study is much smaller than the magnon frequency, which yields an effective dispersive magnon-phonon interaction~\cite{Tang16,Oriol}. We consider the size of the YIG sphere to be much smaller than the MW wavelengths, hence neglecting any radiation pressure on the sphere induced by the MW fields. The Hamiltonian of the system reads
\begin{equation}\label{Hamiltonian}
\begin{split}
{\cal H}/\hbar &= \sum_{j=1,2} \omega_j a_j^{\dag} a_j + \omega_m m^{\dag} m +\frac{\omega_b}{2} (q^2 + p^2) + G_0 m^{\dag} m q \\
&+ \sum_{j=1,2} g_j (a_j^{\dag} m + a_j m^{\dag}) + i \Omega (m^{\dag} e^{-i \omega_0 t} - m e^{i \omega_0 t}),
\end{split}
\end{equation}
where $a_j$ ($m$) and $a_j^{\dag}$ ($m^{\dag}$) are, respectively, the annihilation and creation operators of the cavity mode $j$ (magnon mode), satisfying $[O, O^{\dag}]=1$ $(O=a_j, m)$, and $q$ and $p$ are the dimensionless position and momentum quadratures of the mechanical mode, thus $[q, p]\,\,{=}\,\,i$. $\omega_j$, $\omega_m$, and $\omega_b$ are the resonance frequencies of the cavity mode $j$, the magnon mode, and the mechanical mode, respectively, and the magnon frequency can be adjusted in a large range by altering the external bias magnetic field $H$ via $\omega_m \,{=}\, \gamma_0 H$, where the gyromagnetic ratio $\gamma_0/2\pi \,{=}\, 28$ GHz/T. $G_0$ is the single-magnon magnomechanical coupling rate, and $g_j$ denotes the coupling rate between the magnon mode with the cavity mode $j$, which can be (much) larger than the dissipation rates $\kappa_j$ and $\kappa_m$ of the cavity and magnon modes, $g_j > \kappa_j, \kappa_m$, leading to cavity-magnon polaritons~\cite{Strong1,S2,S3,S4,S5,S6,S7,S8}. The Rabi frequency $\Omega=\frac{\sqrt{5}}{4} \gamma_0 \sqrt{N} B_0$~\cite{JiePRL} denotes the coupling strength between the magnon mode and its driving magnetic field with frequency $\omega_0$ and amplitude $B_0$, where the total number of spins $N \,{=}\, \rho V$ with the spin density of YIG $\rho=4.22 \times 10^{27}$ m$^{-3}$ and the volume of the sphere $V$.

For convenience, we switch to the rotating frame with respect to the drive frequency $\omega_0$, and by including input noises and dissipations of the system, we obtain the following quantum Langevin equations (QLEs)
\begin{equation}\label{QLES}
\begin{split}
\dot{a_j}&= - i \Delta_j a_j - i g_j m - \kappa_j a_j + \sqrt{2 \kappa_j} a_j^{\rm in},  \,\,\,\,\,\, (j{=}1,2)  \\
\dot{m}&= - i \Delta_m m - i \sum_{j=1,2} g_j a_j -i G_0 m q + \Omega - \kappa_m m + \sqrt{2 \kappa_m} m^{\rm in},  \\
\dot{q}&= \omega_b p,   \\
\dot{p}&= - \omega_b q - G_0 m^{\dag}m - \gamma p  + \xi,  \\
\end{split}
\end{equation}
where $\Delta_j \,\,{=}\,\, \omega_j - \omega_0$, $\Delta_m = \omega_m - \omega_0$, $\gamma$ is the mechanical damping rate, and $a_j^{\rm in}$, $m^{\rm in}$ are input noise operators with zero mean value acting on the cavity and magnon modes, respectively, which are characterized by the following correlation functions~\cite{Zoller}: $\langle a_j^{\rm in}(t) \, a_j^{\rm in \dag}(t')\rangle \,\,{=}\,\, \big[ N_j(\omega_j)\, {+}\,1 \big] \,\delta(t\,{-}\,t')$, $\langle a_j^{\rm in \dag}(t) \, a_j^{\rm in}(t')\rangle \,\,{=}\,\, N_j(\omega_j) \, \delta(t\,{-}\,t')$, and $\langle m^{\rm in}(t) \,\, m^{\rm in \dag}(t')\rangle = \big[ N_m (\omega_m) \,\,{+}\,1 \big] \, \delta(t\,{-}\,t')$, $\langle m^{\rm in \dag}(t) \, m^{\rm in}(t')\rangle \,\,{=}\,\, N_m(\omega_m)\,\, \delta(t\,{-}\,t')$. The Langevin force operator $\xi$, accounting for the Brownian motion of the mechanical oscillator, is autocorrelated as $\langle \xi(t)\xi(t')\,\,{+}\,\,\xi(t') \xi(t) \rangle/2 \,\,\, {\simeq} \,\,\, \gamma \,\, \big[ 2 N_b(\omega_b) \,\,{+}\,\, 1 \big] \,\, \delta(t\,{-}\,t')$, where a Markovian approximation has been taken valid for a large mechanical quality factor $Q_m = \omega_b/\gamma \gg 1$~\cite{Markov}. The equilibrium mean thermal photon, magnon, and phonon numbers are $N_k(\omega_k) = \Big[{\rm exp} \Big( \frac{\hbar \omega_k}{k_B T} \Big)-1 \Big]^{-1}$ ($k = 1,2,m,b$), with $k_B$ the Boltzmann constant and $T$ the environmental temperature.

Because the magnon mode is strongly driven, it has a large amplitude $|\langle m \rangle| \gg 1$, and further owing to the cavity-magnon beamsplitter interactions the two cavity fields are also of large amplitudes. This allows us to linearize the system dynamics around semiclassical averages by writing any mode operator as a c-number plus its fluctuation operator $O=\langle O \rangle +\delta O$, ($O\, {=}\, a_j,m,q,p$), and neglecting small second-order fluctuation terms. Substituting those linearized mode operators into Eq.~\eqref{QLES}, the equations are then separated into two sets of equations, respectively, for semiclassical averages and for quantum fluctuations. The solutions of the averages are obtained, which are $\langle p \rangle = 0$, $\langle q \rangle = - \frac{G_0}{\omega_b} |\langle m \rangle|^2$, $\langle a_j \rangle =  \frac{-i g_j}{i \Delta_j + \kappa_j} \langle m \rangle$, and $\langle m \rangle$ is given by
\begin{equation} \label{Mag.Ampli.}
\begin{split}
&\langle m \rangle = \\
&\frac{\Omega (i \Delta_1 + \kappa_1)(i \Delta_2 + \kappa_2)}{(i \tilde{\Delta}_m {+} \kappa_m)(i \Delta_1 {+} \kappa_1)(i \Delta_2 {+} \kappa_2) + g_1^2(i \Delta_2 {+} \kappa_2) + g_2^2(i \Delta_1 {+} \kappa_1)},
\end{split}
\end{equation}
with $\tilde{\Delta}_m = \Delta_m + G_0 \langle q \rangle$ the effective detuning of the magnon mode including the frequency shift caused by the magnetostrictive interaction. It takes a simpler form
\begin{equation}\label{SimpleForm}
\langle m \rangle \simeq  \frac{i \Omega \Delta_1 \Delta_2}{- \tilde{\Delta}_m \Delta_1 \Delta_2 + g_1^2 \Delta_2 + g_2^2 \Delta_1},
\end{equation}
when $|\Delta_j|, |\tilde{ \Delta}_m| \gg  \kappa_j, \kappa_m $. Let us introduce the quadratures of the quantum fluctuations $(\delta X_1, \delta Y_1, \delta X_2, \delta Y_2, \delta x, \delta y, \delta q, \delta p)$, where $\delta X_j=(\delta a_j + \delta a_j^{\dag})/\sqrt{2}$, $\delta Y_j=i(\delta a_j^{\dag} - \delta a_j)/\sqrt{2}$, $\delta x=(\delta m + \delta m^{\dag})/\sqrt{2}$, and $\delta y=i(\delta m^{\dag} - \delta m)/\sqrt{2}$, and the input noise quadratures are defined in the same way. The QLEs of the quadrature fluctuations can be cast in the matrix form
\begin{equation}
\dot{u} (t) = A u(t) + n(t) ,
\end{equation}
where $u(t){=}\big[\delta X_1 (t),\! \delta Y_1 (t),\! \delta X_2 (t),\! \delta Y_2 (t),\! \delta x (t),\! \delta y (t),\! \delta q (t),\! \delta p (t) \big]^T$, $n (t) {=} \big[ \!\sqrt{2\kappa_1} X_1^{\rm in} (t), \! \sqrt{2\kappa_1} Y_1^{\rm in} (t), \! \sqrt{2\kappa_2} X_2^{\rm in} (t), \! \sqrt{2\kappa_2} Y_2^{\rm in} (t), \! \sqrt{2\kappa_m} x^{\rm in} (t), \\  \! \sqrt{2\kappa_m} y^{\rm in} (t), 0, \xi (t) \big]^T$ is the vector of noises entering the system, and the drift matrix $A$ is given by
\begin{equation}\label{AAA}
A =
\begin{pmatrix}
-\kappa_1  & \, \Delta_1  \, &  0 &  0 &  0 &  g_1  &  0  &  0   \\
-\Delta_1  & \, -\kappa_1  \, & 0  & 0 & -g_1  & 0  &  0  &  0   \\
0 & 0 & -\kappa_2  & \Delta_2 & 0  &  g_2  & 0  &  0 \\
0 & 0 & -\Delta_2 & -\kappa_2 & -g_2  &  0  &  0  &  0 \\
0 & g_1  & 0  &  g_2  & -\kappa_m  &  \tilde{\Delta}_m &  -G  &  0 \\
-g_1  & 0 & -g_2  &  0  &  -\tilde{\Delta}_m & -\kappa_m & 0  &  0 \\
0 &  0  &  0  &  0  &  0  & 0  &  0  &  \omega_b   \\
0 &  0  &  0  &  0  & 0  &  G  & -\omega_b & -\gamma   \\
\end{pmatrix} ,
\end{equation}
where $G = i \sqrt{2} G_0 \langle m \rangle$ is the effective magnomechanical coupling rate. By using the result of Eq.~\eqref{SimpleForm}, we obtain
\begin{equation}\label{GSF}
G \simeq \frac{\sqrt{2} G_0 \Omega \Delta_1 \Delta_2}{\tilde{\Delta}_m \Delta_1 \Delta_2 - g_1^2 \Delta_2 - g_2^2 \Delta_1},
\end{equation}
which shows that the coupling can be significantly enhanced with a strong magnon drive.

We are interested in the quantum correlation of two MW fields in the steady state. The steady state of the system is a four-mode Gaussian state due to the linearized dynamics and the Gaussian nature of input noises. Such a state is fully characterized by an $8 \times 8$ covariance matrix (CM) ${\CC}$ with its entries defined as ${\CC}_{ij}(t) = \frac{1}{2} \langle u_i(t) u_j(t') + u_j(t') u_i(t) \rangle$ $(i,j = 1,2,...,8)$. It can be obtained straightforwardly by solving the Lyapunov equation~\cite{Hahn}
\begin{equation}\label{Lyap}
A \CC+\CC A^T = -\DD,
\end{equation}
where $\DD={\rm diag} \big[ \kappa_1 (2N_1+1), \kappa_1 (2N_1+1), \kappa_2 (2N_2+1),  \kappa_2 (2N_2+1), \kappa_m (2N_m{+}1),  \kappa_m (2N_m{+}1), 0,  \gamma (2N_b {+}1 ) \big]$ is the diffusion matrix, whose entries are defined through $\langle  n_i(t) n_j(t') +n_j(t') n_i(t) \rangle/2 = \DD_{ij} \delta (t-t')$.

We adopt the logarithmic negativity~\cite{LogNeg} to quantify the entanglement between the two MW cavity fields. It is a full entanglement monotone under local operations and classical communication~\cite{Plenio} and an upper bound for the distillable entanglement~\cite{LogNeg}. The logarithmic negativity for Gaussian states is defined as~\cite{Adesso}
\begin{equation}
E_N := \max \big[ 0, \, -\ln2\tilde\nu_- \big] ,
\end{equation}
where $\tilde\nu_-\,\,{=}\,\min{\rm eig}|i\Omega_2\tilde{\CC}_{mw}|$ (with the symplectic matrix $\Omega_2=\oplus^2_{j=1} \! i\sigma_y$ and the $y$-Pauli matrix $\sigma_y$) is the minimum symplectic eigenvalue of the CM $\tilde{\CC}_{mw}=\PP \CC_{mw} \PP$, with $\CC_{mw}$ the CM of the two MW fields, which is obtained by removing in $\CC$ the rows and columns associated with the magnon and mechanical modes, and $\PP={\rm diag}(1,-1,1,1)$ is the matrix that performs partial transposition on CMs~\cite{Simon}.

\begin{figure}[t]
\includegraphics[width=9.0cm,clip]{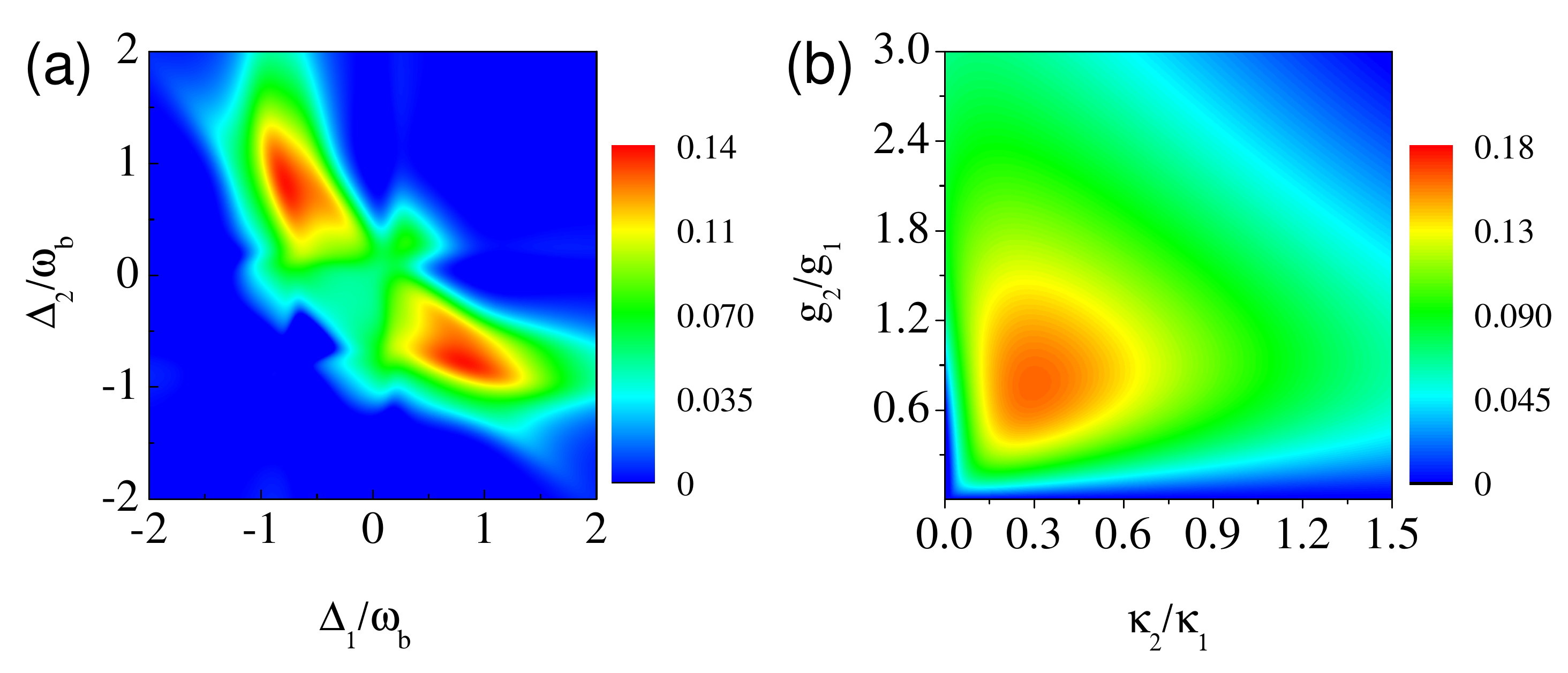}
\caption{(a) Density plot of the entanglement $E_N$ between two MW cavity fields vs (a) $\Delta_1$ and $\Delta_2$, (b) $\kappa_{2}/\kappa_{1}$ and $g_{2}/g_{1}$ ($\kappa_{1}$, $g_{1}$ are fixed). We take $\tilde{\Delta}_m \,\,{=}\,\, 0.9 \omega_b$, $\kappa_2 \,\,{=}\,\, \kappa_1$, $g_2 \,\,{=}\,\, g_1$ in (a), and $\Delta_1 = -\Delta_2 = \omega_b$ in (b). See text for the other parameters.}
\label{Fig2}
\end{figure}

\begin{figure}[b]
\includegraphics[width=\linewidth]{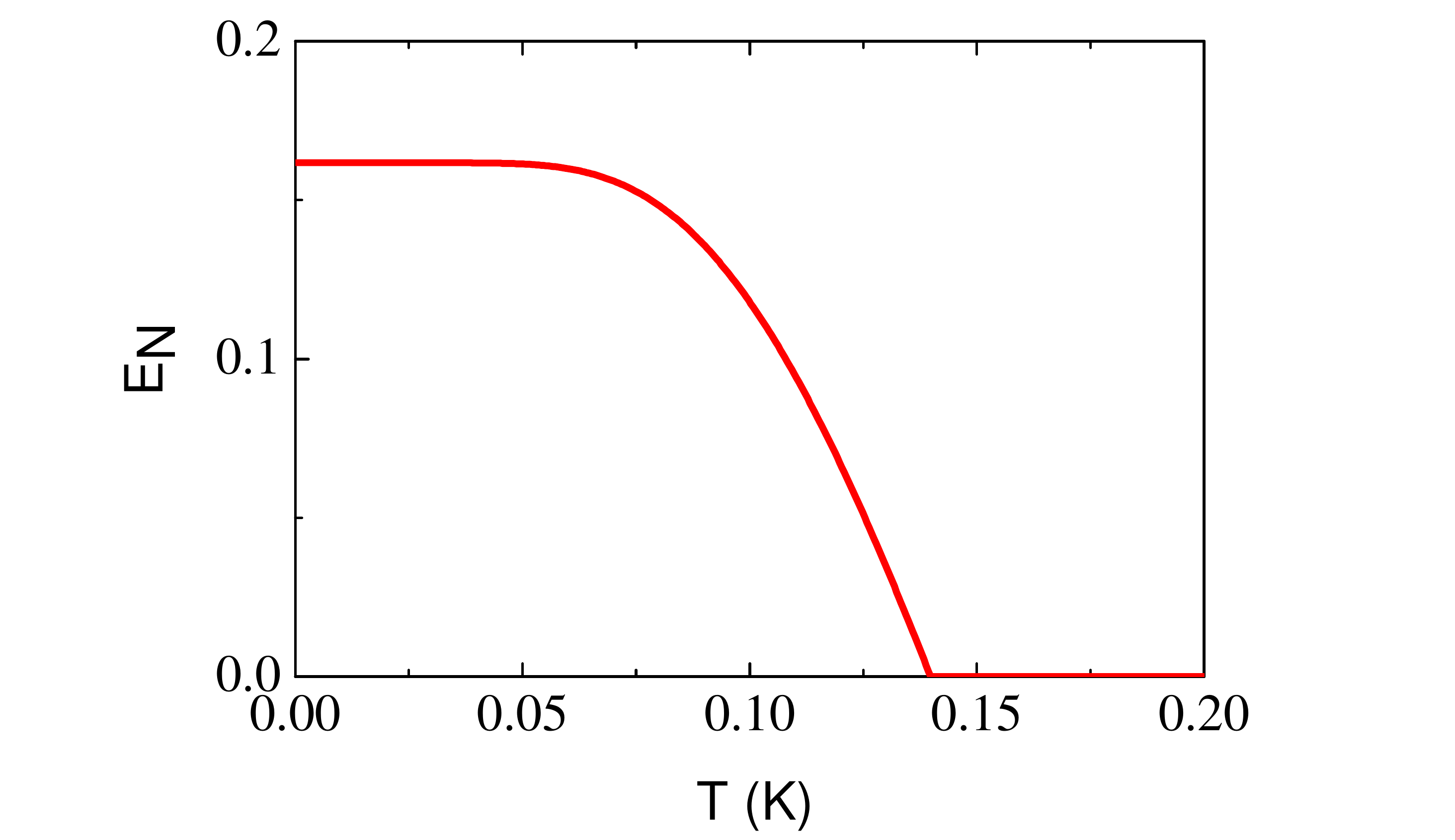}
\caption{MW entanglement $E_N$ vs temperature $T$. The parameters are those with which the maximum entanglement is achieved in Fig.~\ref{Fig2}(b). }
\label{Fig3}
\end{figure}

{\it MW entanglement and its detection}. In Fig.~\ref{Fig2} we present the main results of the entanglement between two MW cavity fields. The stationary entanglement is guaranteed by the negative eigenvalues (real parts) of the drift matrix $A$. Figure~\ref{Fig2}(a) shows clearly that the maximum entanglement is achieved when the two cavity fields are respectively resonant with the two mechanical sidebands [see Fig.~\ref{Fig1}(b)], i.e., $\Delta_1=-\Delta_2 \simeq \pm \, \omega_b$, where ``$\pm$" sign is taken due to the symmetry of the two cavity fields. And the magnon mode resonant with the blue sideband $\tilde{\Delta}_m \simeq \omega_b$ corresponds to the anti-Stokes process, which significantly cools the phonon mode, thus eliminating the main obstacle for observing entanglement~\cite{JiePRL}. We have employed experimentally feasible parameters~\cite{Tang16}: $\omega_m/2\pi \,\,{=}\,10$ GHz, $\omega_b/2\pi \,\,{=}\,10$ MHz, $\gamma/2\pi \,\,{=}\,\,10^2$ Hz, $\kappa_m/2\pi\,\,{=}\,\kappa_1/2\pi\,\,{=}\,\,1$ MHz, $g_{1}/2\pi \,\,{=}\,\, 3.8$ MHz, $G/2\pi \,\,{=}\,\,4.5$ MHz, and $T\,{=}\,20$ mK. We use a strong magnon-phonon coupling $G > \kappa_m$ to create magnomechanical entanglement. This means that a strong magnon driving field should be used, and in order to avoid unwanted magnon Kerr effect~\cite{You18,Zhedong} the bare coupling rate $G_0$ should not be too small~\cite{JiePRL,JieRC}. For the optimal case $\Delta_1=-\Delta_2 \simeq \pm \, \omega_b$ in Fig.~\ref{Fig2}(a), a driving power of 6.3 mW (0.57 mW) should be used to yield $G/2\pi \,\,{=}\,\,4.5$ MHz for $G_0/2\pi \,\,{=}\,\,0.3$ Hz (1 Hz), while keeping the Kerr effect negligible. In Fig.~\ref{Fig2}(b), we analyse the optimal coupling rates $g_{1,2}$ and decay rates $\kappa_{1,2}$, and find that in both situations $\Delta_1=-\Delta_2 \simeq \omega_b$ and $-\omega_b$, close coupling rates should be used, and the cavity that is resonant with the red (blue) mechanical sideband should have a smaller (larger) decay rate than the other. Such an asymmetric feature is due to the different roles of the two sidebands. The entanglement is robust against environmental temperature and survives up to $\sim$140 mK, as shown in Fig.~\ref{Fig3}, below which the average phonon number is always smaller than 1, showing that mechanical cooling is thus a precondition to observe quantum entanglement in the system~\cite{JiePRL}. The generated MW entanglement can be detected by measuring the CM of two cavity output fields. Such measurement in the MW domain has been realized in the experiments~\cite{Lehnert,SB}.

\begin{figure}[t]
\includegraphics[width=9.0cm,clip]{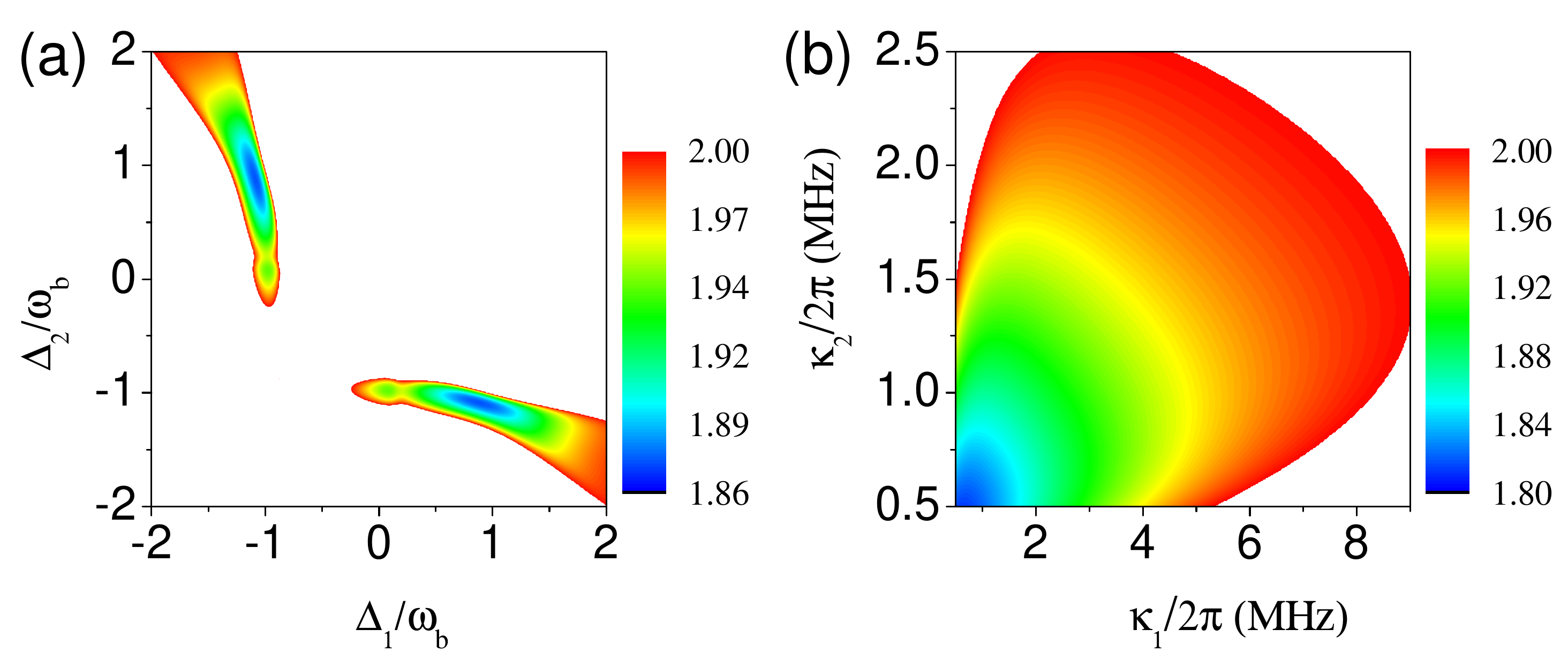}
\caption{Density plot of $\langle \delta X_+^{2} \rangle + \langle \delta Y_-^{2} \rangle$ vs (a) $\Delta_1$ and $\Delta_2$, (b) $\kappa_1$ and $\kappa_2$. The blank areas denote $\langle \delta X_+^{2} \rangle + \langle \delta Y_-^{2} \rangle >2$. The parameters are the same as in Fig.~\ref{Fig2}(a), and we take optimal detunings $\Delta_1=0.9 \omega_b$ and $\Delta_2= -1.1 \omega_b$ in (b). }
\label{Fig4}
\end{figure}

Alternatively, one can also {\it verify} the entanglement by using the Duan criterion~\cite{Duan}, which requires simpler experimental operations, i.e., one does not have to measure all the entries of the $4 \times 4$ CM, but measure only two collective quadratures~\cite{Finland}. Specifically, a sufficient condition for entanglement is that the two collective quadratures satisfy the inequality  
\begin{equation}
\langle \delta X_+^{2} \rangle + \langle \delta Y_-^{2} \rangle < 2,
\end{equation}
where $X_+ = X_1 +X_2$, and $Y_- = Y_1 -Y_2$. Figure~\ref{Fig4}(a) shows that in two areas around $\Delta_1=-\Delta_2 \simeq \pm \omega_b$ the inequality is fulfilled, indicating that the two cavity fields are entangled.

\begin{figure}[b]
\includegraphics[width=\linewidth]{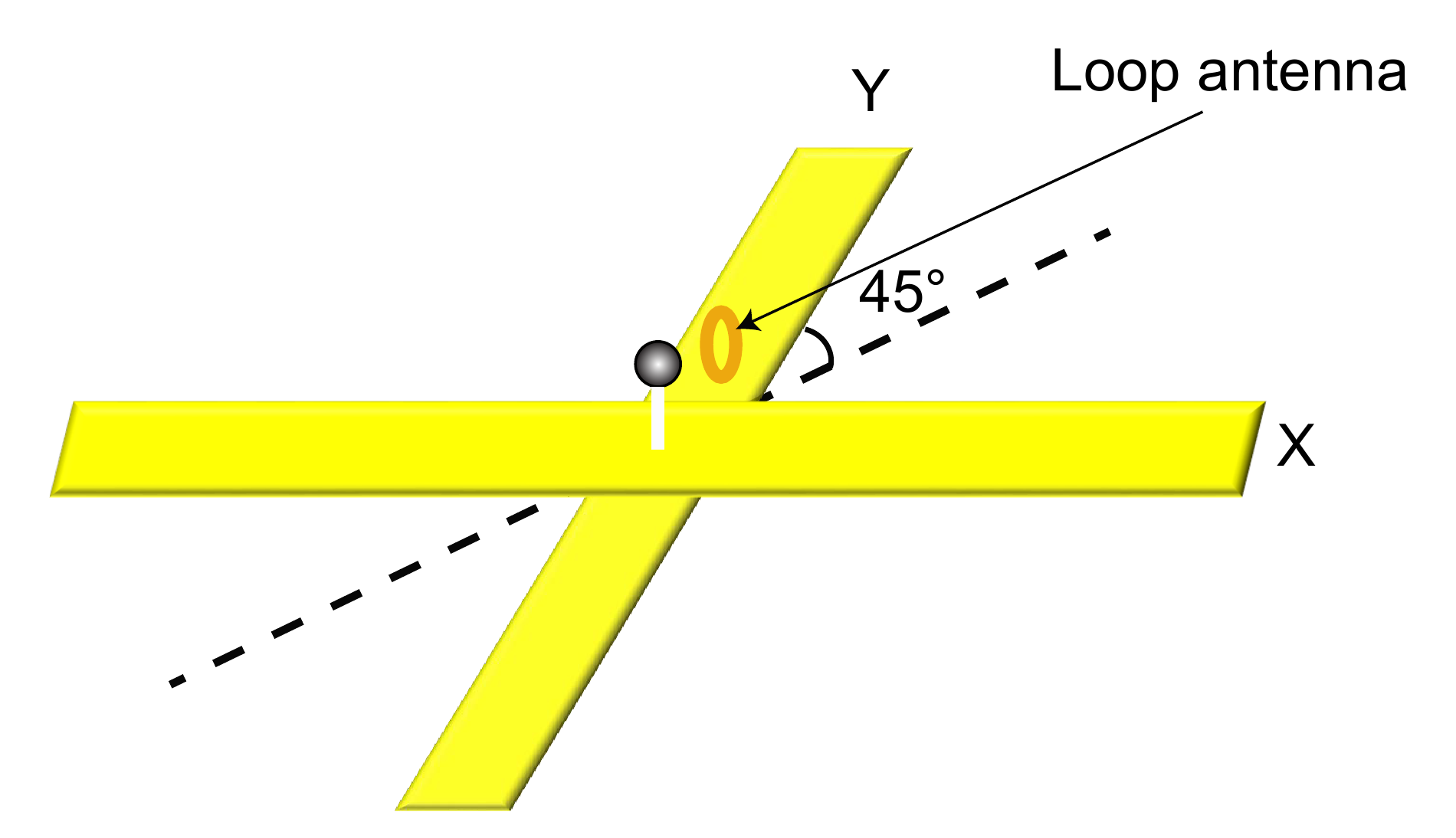}
\caption{ A YIG sphere is placed near the maximum magnetic fields of two MW cavity fields in a cross-shape cavity. A loop antenna at the end of a superconducting MW line is used to drive the magnon mode~\cite{You18}. }
\label{Fig5}
\end{figure}

{\it Experimental implementations}. We now discuss possible configurations that could realize the proposal. Two MW cavities and each cavity containing a cavity mode are preferred. In this situation, the frequencies of the cavity fields can be adjusted flexibly to match the two mechanical sidebands. The two cavities could be placed perpendicularly in the horizontal plane with the YIG sphere located in the intersection (near the maximum magnetic fields) of the cavity fields. This can be realized in a planar cross-shape cavity~\cite{HuNC} or coplanar waveguide~\cite{coplanar}, see Fig.~\ref{Fig5}. Taking the ``X"-cavity~\cite{HuNC} as an example, one can set the bias magnetic field along the $z$ (vertical) direction, the magnetic fields of two cavity modes along the $x$ and $y$ direction, respectively, and the driving magnetic field in the $x$-$y$ plane and of e.g., 45 degrees with both the $x$ and $y$ direction. For directly driving the magnon mode, one may adopt a superconducting MW line with a small loop antenna at its end~\cite{You18}. In this case, the loop antenna will also couple to the two cavity modes leading to increased cavity decay rates. However, owing to its relatively small dimension compared with the cavity setup the influence is only moderate~\cite{Yipu}. Besides, the cross configuration of the cavity may also reduce the $Q$ factor of the cavities induced by the damage to boundary conditions. Taking into account the aforementioned effects, we study the Duan criterion for taking larger cavity decay rates in Fig.~\ref{Fig4}(b). It shows that with much larger decay rates, the two cavity fields are still entangled. Given the flexibility of the cavity resonant frequencies, the mechanical frequency can be freely chosen in a large range (always keeping it much smaller than the magnon frequency). The results presented in this work employed a $\sim$10 MHz mechanical mode of a 250-$\mu$m-diameter YIG sphere~\cite{Tang16}. For such a large sphere, the bare magnomechanical coupling is small, but it can be increased by using a smaller sphere such that the pump power required is reduced, which can weaken both the unwanted nonlinear effect and the by-effect of the coupling of the loop antenna to the cavity modes.


{\it Conclusions.} We present a new mechanism for creating MW entangled states based on magnetostrictive interaction in a ferrimagnetic YIG sphere. The mechanism makes use of the nonlinearity of such a magnomechanical interaction. The entanglement is in the steady state and robust against cavity dissipations and environmental temperature. We show strategies to detect the entanglement and a possible configuration that is promising to realize the proposal. We analyse in detail various practical imperfections which would help future experimental realizations. This work may find applications in quantum information science, quantum metrology, and quantum tasks that require entangled CV MW fields.

{\it Acknowledgments}. We thank Junjie Liu and Yi-Pu Wang for fruitful discussions on potential experimental realizations. This work has been supported by the National Key Research and Development Program of China (Grants No. 2017YFA0304200 and No. 2017YFA0304202), the Royal Society Newton International Fellowship (NF170876) of UK, and the European Research Council project (ERC StG Strong-Q, 676842).

\end{document}